# Dynamical Systems, Representation of Particle Beams

*Alex Chao*
*SLAC National Accelerator Laboratory, Menlo Park, California, USA*

**Abstract**
An overview of dynamical systems in accelerator physics is presented with a suggestion of a few issues to be addressed. Also mentioned are a few possible developments in the future. Technical details supporting the views are not presented.

**Keywords**
Dynamical systems; phase space; Liouville theorem; single-particle models; multi-particle models; binary models; beam-beam effects; dynamic aperture; detuned integrable systems; microbunches.

## 1. Introduction

In the book "One Two Three…. Infinity", George Gamow [1] asked a question:

> "How high can you count?"

One can imagine that a response from an audience might take the following form:

> "One!" was the first answer.
> "Two!" was a more sophisticated answer.
> "Three!" came an advanced answer after a pause.
> "Any higher?"

After a long pause, the audience replied:

> "We give up. Any number larger is so confusing. Let's call it Infinity!"

There are no numbers of significance between three and infinity.

Who could be the audience? A primitive people? Maybe. But accelerator physicists are no better.

Basically, the two big areas in accelerator beam dynamics are

   i)  single particle dynamics, and
   ii) collective effects.

The single particle dynamics is a one-particle model. The number of particles is "One!" Collective effects are treated assuming the beam is a continuum. The number of particles is "Infinity!"

We basically have not much in between, while the actual beam has, e.g., $10^{12}$ particles. We have therefore the dilemma, how do we represent and analyse the actual dynamical system, get results, and then trust the results? The dilemma is illustrated in Fig. 1.

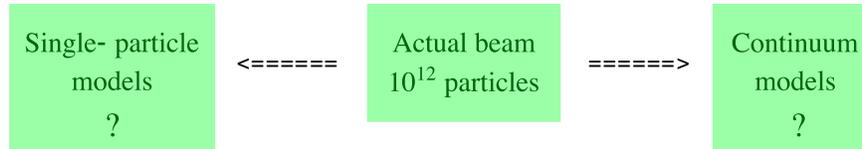

**Fig. 1:** The actual beam contains $10^{12}$ particles. Do we model the actual system as a single particle, or do we model it as a continuum? In other words, is $10^{12}$ closer to 1 or closer to ∞?

In the following, I shall mention some jargon and make a few comments. The discussions are not in a particular order or for a particular purpose, but hopefully some comments will become useful along the way. The jargon we discuss includes:

i) Single-particle models – 1;
ii) Two-particle models – 2;
iii) Continuum models – ∞;
iv) Multiparticle models – $10^6$;
v) Dynamic apertures;
vi) Microbunches.

## 2. Single-particle models – a recap of what you already learned

In the single-particle models, the beam is represented by 'point particles', particles with no internal structure, no size, but yet having mass and charge, sometimes even spin. That means we are suggesting an object that has no size and yet carries an angular momentum—speak of a contradiction in terms!

There are at least three reasons why this concept of the point particle can only be flawed:
 i) It is inconceivable that something has a mass and a spin but no size.
 ii) It violates the uncertainty principle of quantum mechanics.
 iii) It leads to fierce divergences when the particle also has a charge – the idea of a 'point charge' is even worse than the idea of a 'point particle'.
But we accelerator physicists sweep these problems under the rug (as we will do in the following discussion).

Swallowing the concepts of point particles and point charges then, the beam is represented by a collection ($10^{12}$ of them) of single particles. As long as they do not interact among themselves, these models describe the beam faithfully simply by repeating the analysis $10^{12}$ times.

The single-particle models have been very successful, yielding deep knowledge as elucidated by many of the other lectures at this school.

Where does the success come from? The success has been based on the support of the Liouville theorem as its backbone. Perhaps I should explain.

To appreciate the intricacy of the single-particle models, one needs to appreciate the phase space. The Liouville theorem states that all those intricate subtle dynamical effects in phase space, predicted by the single-particle models, are rigorously preserved, and therefore last a surprisingly long time, i.e., forever. All predictions by single-particle models, including the most intricate effects down to the finest details, happen without dilution, and that in turn provides the root of the success of the single-particle models. Note that all these intricacies occur in the 6D phase space, while they are clueless in the 3D real space.

The combination of the phase space plus the Liouville theorem is the basis of a very large number of accelerator applications:
   i) Courant–Snyder dynamics;
   ii) emittance preservation;
   iii) RF gymnastics;
   iv) phase space displacement acceleration;
   v) KAM theorem [2-4];
   vi) 6D phase space gymnastics;
   vii) emittance exchanges; echoes;
   viii)     echoes;
   ix) free electron lasers;
   x) harmonic generation techniques;
   xi) steady state microbunching techniques
             etc.

This is an awesome list. We have indeed come a very long way with the single-particle models.

## 3. A note on history

In the 1960s and 70s, accelerator nonlinear dynamics was done by Hamiltonian dynamics and the canonical perturbation theories. There had been some important accomplishments:

- Single-resonance analyses was a great success.
- But they break down with multiple resonances and divergences by small denominators.
- Most advanced tool was COSY [5] 5th order by integrating the Equation of Motion (EOM)
- Cumbersome to proceed, continued pursuit along this line seemed stuck.

In the 1980s, two breakthroughs revolutionized the landscape:

- Lie algebra by Alex Dragt [6].
- Truncated Power Series Algebra (TPSA) by Martin Berz [7, 8].
- These two efforts combined forces and blossomed at the SSC, driven by the SSC need at the SSC Central Design Group.

Today, the canonical perturbation theories are gone. The Lie framework plus the TPSA technique is the powerful industry standard for single-particle dynamics. TPSA provides the most efficient way to obtain maps, superseding all previous techniques. Assuming convergence (ignore chaos), Lie algebra then extracts from TPSA to obtain the analysis of the one-turn effective Hamiltonian and normal forms.

TPSA and Lie algebra are most powerful if and only if used together. It is suggested that research and learning be performed with their hand-in-hand combination in mind. Learning only one side without the other is considered highly nonoptimal.

But this revolution was >30 years ago! Perhaps we should not call it a 'modern approach' anymore. To put in things in context, the canonical perturbative theories have lived only 20 years in their entire lives!

But have we completed the revolution? Has the old Hamiltonian dynamics gone completely? Are we done? The answer is no! Clear evidence is that the very basic old structure has remained, as evidenced by the stubborn Courant–Snyder language of the beta-functions and their family of special functions.

The formalism based on the beta-functions, dispersion functions, *H*-functions, etc. is a remnant of the now extinct perturbative Hamiltonian dynamics and is inconsistent with the TPSA. The correct use of TPSA necessarily abandons the use of these special functions – and it is worth noting that discussions of these special functions occupy half of every accelerator physics textbook!

An alternative framework developed in 1979, now 40 years ago, was called SLIM [9]. It advocates no use of these special functions and it is consistent with TPSA to a first-order linear analysis. Use of SLIM and TPSA will shorten the standard accelerator physics textbooks by half.

## 4. Two-particle models

Setting aside quantum mechanics, single-particle models are a tremendous success. A weakness occurs when particles interact electromagnetically with each other or with the vacuum chamber environment. The field a particle sees is then no longer prescribed by the external field alone. Pursuing multiple particles along this line will immediately become impossible as the number of particles is increased.

Some progress can be made if we have only two particles in the beam. The two particles interact with the environment or with each other. We thus arrive with the two-particle models.

When the two point particles interact with each other, the analysis developed in single-particle models can still be applied but it becomes cumbersome. So far, we have only results in simplified models that are nowhere near the sophistication of the single-particle models—no phase space manipulation, no KAM theorem [2-4] applications (at least not yet). On the other hand, with only two particles, and with sufficiently drastic simplifications, these two-particle models can be solved analytically.

When solved, these two-particle models yield important insights towards the unravelling of collective effects. We shall leave out this discussion below. Instead, let us mention another insightful consequence of the two-particle models, i.e., here we found two approaches to describe the particles' dynamics.

1. We can consider the motion of $x_1(t)$ and $x_2(t)$ as two individual point particles evolving in time.

2. Or we can describe it as a superposition of two 'modes', a + mode in which the two particles move together *(x₁(t) + x₂(t))*, and a – mode with the two particles move oppositely to each other *(x₁(t) - x₂(t))*.

These are two representations of the beam dynamics, the 'particle representation' and the 'mode representation'.

The two representations are completely equivalent. They necessarily yield exactly the same final results. The particle representation is also called a 'time-domain' approach. The mode representation is also called a 'frequency-domain' approach. Again, these two approaches necessarily give identical final results.

It is called a time domain because in the single-particle representation, we focus on the time evolution of the two particles $x_1(t)$ and $x_2(t)$. It is called a frequency domain in the mode representation because we focus on the eigen-frequencies of the two modes. The two representations are preferred for use by different people. Simulation programs might prefer the time domain (computer). Beam stability analysis might bias toward the frequency domain (pencil and paper). Neither deserves to claim advantage over the other.

We see that single-particle models are exclusively using the time domain. As we shall see later, continuum analysis uses exclusively the frequency domain. The two prime areas of accelerator physics use two completely opposite beam representations. The two-particle models serve as an intermediate, a geometric mean, of these two pictures.

## 5. Continuum models

The opposite extreme to single-particle models is the continuum models. The beam is now represented by a continuum of distribution in the 6D phase space (note: not only in the real 3D space). All discreteness of point particles is smoothed out. Beam evolution is determined by the Vlasov equation.

Without particles, the analysis of beam motion is now described as a superposition of 'modes'. Choosing a finite number of particles in a time-domain computer simulation is then equivalent to truncation to the highest mode number in a frequency-domain analytical calculation. Table 1 gives a brief comparison of the single-particle and the continuum models.

Table 1: A brief comparison between the single-particle and the continuum models

|  | Single-particle dynamics | Collective effects |
|---|---|---|
| Number of particles | 1 | ∞ |
| Beam representation | Collection of point particles | Continuum in phase space distribution |
| Dynamics approach | Time domain | Frequency domain |
| Analysis | $x(t)$ | modes |

## 6. Binary models

So we have tools developed for a single particle, two particles, and a continuum distribution of particles. We let go of the attempt to represent $10_{12}$ particles. In doing so, have we missed important effects? Should we be concerned?

For example, in the continuum models, a drastic approximation is being made. In the Vlasov approach, we ignore collisions among point particles, i.e., we ignore interacting fields between individual particles. And yet, the collective field of the continuum beam has been included. In other words, we only recognize the averaged sum of the fields from all particles and ignore the highly fluctuating fields of the individual particles.

This approximation might work for the collective effects due to vacuum-chamber wakefields. It is however a drastic assumption for space charge effects recognizing the fact that the Coulomb field diverges between individual point particles and therefore has to be extremely singular and granular, while the sum of fields from a continuum beam is smooth. Perhaps the question can be asked: Do we really believe we can replace the force in Fig. 2(a) by that of Fig. 2(b)?

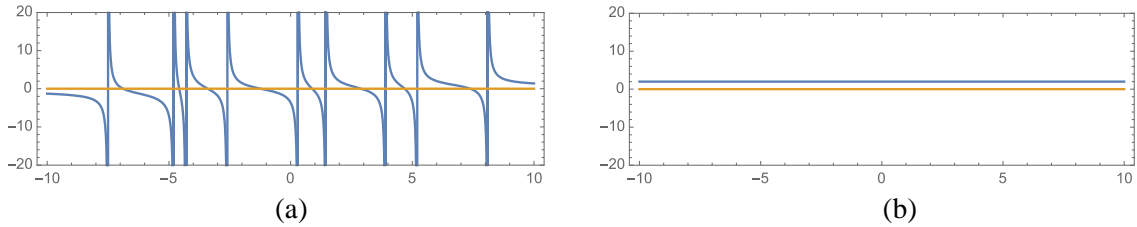

(a)          (b)

**Fig. 2:** (a) A sketch of what a space charge force actually looks like. (b) A smoothed force that we use to calculate space charge effects. Can we really trust the results we get using model (b)?

The way we have tried to deal with the problem is to insist on considering smoothed collective effects, but then to supplement it by a few binary models such as intrabeam and Touschek modifications. This is a partial remedy. Clearly it is not complete, but is it sufficient?

## 7. Beam–beam models

The situation of smoothing out a granular force by a smooth force gets worse with the beam–beam effects, as Fig. 3 intends to illustrate. Which picture should be used to model the oncoming beam when evaluating the beam–beam effects?

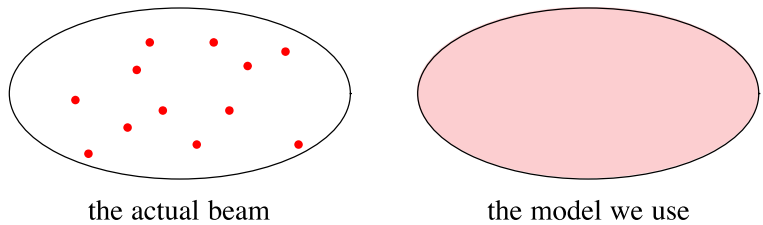

the actual beam          the model we use

**Fig. 3:** A comparison between an actual beam and a model beam we use to calculate the beam–beam effects.

Is the beam–beam limit determined by the smooth kick (as we have been doing all along)? Or is part of it actually due to the granularity of the oncoming beam distribution? Have we left out some important noise-like diffusion effect? At least as a partial remedy, should we develop a beam–beam binary collision model like we did for Toushek and intrabeam?

## 8. Multiparticle models, simulations, macroparticles

Back to the single-particle models. The obvious next step is to add more particles in the beam representation. However, when there are 3 or more particles, the analysis becomes cumbersome even for over-simplified models as we did for the two-particle models.

So far, the only analytically accessible case without computer simulations are:

- single-particle linear systems;
- over-simplified linearized two-particle models;
- linearized continuum models.

For all other cases, we use computer simulations.

Note that two-particle models were the first step to increasing the number of particles. It is just that they are simple enough that we can still deal with them analytically sometimes. When the number of particles is increased to three or more, we then evoke computer simulations. We can of course do two-particle models using computers as well.

To study multiparticle collective effects by simulations, we have two ways to proceed.

- Represent the beam by a collection of single point particles that interact with wakefields.
- Divide the otherwise continuum phase space into grids, each grid represented by a 'macroparticle'.

Both ways are time-domain approaches, but they are not quite the same concept. One describes the motion of particles, the other describes the time evolution of the elements in phase space. The reason for their one-to-one correspondence is attributed to the Liouville theorem.

In treating multiparticle collective effects, time-domain applications use 'wakefields'. The frequency domain uses 'impedances'. Wakefields are more used in simulations. Impedances are more used in analyses. Table 2 gives a comparison. These multiparticle models typically go up to $10^6$ particles, or $10^6$ modes. On the other hand, it is stated here that there is basically no difference between $10^6$ and Gamow's 'three!'

**Table 2:** A comparison between time-domain and frequency-domain treatments of collective effects

| Simulations | Analyses |
|---|---|
| Time domain | Frequency domain |
| Wakefields | Impedances |

## 9. Should we try to simulate $10^{12}$ particles?

Assuming computer power is available, should we try to 'do it right' and simulate the case with $10^{12}$ particles? To address this question, we should first ask the question: what for? Here is the situation:

- Single-particle models allow exploration of detailed phase space without collective effects.
- Two-particle models give the qualitative understanding of collective effects.
- A $10^6$ particle computer simulation gives sufficiently accurate information on lower-order modes and instability thresholds.
- What is there to learn from a $10^{12}$ particle simulation?

## 10. Dynamic aperture

Dynamic aperture is an old problem that is intrinsically difficult because the system is not integrable. All 'soluble' cases <u>assume</u> a priori integrability, e.g.:

  i) single isolated resonance; or
  ii) convergence in power series expansions, either Lie algebra or TPSA.

However, this is a solution by assuming a solution. The question remains: does TPSA converge? To what order can we truncate TPSA? There might never be an answer to this question. Our approach can only be: let us assume the convergence and solubility, and push to the limit to see how it breaks down!

We have no handle on overlapping resonances, or the chaos/nonintegrable cases. For those cases, the only dependable tool has been computer simulations. Or is it?

We need a sizable region of stability (in phase space) to situate the beam in a storage ring. KAM theorem, however, does not give any sizable stability region for the beam at all. KAM mathematicians are terrified to find that we require such a huge region for stability. And yet storage rings do work!

The catch lies in the fact that the mathematicians, the KAM, address stability for infinite number of turns. But we ask for stability for a mere $10^{10}$ turns. These are completely different issues.

Note that the Earth has evolved around the Sun only for a few $10^9$ turns. KAM mathematicians would place grave doubt on the stability of the Earth, and they in fact rightly do.

Extensive efforts were made for example in the dynamic aperture studies for the Superconducting Super Collider. One of the results is shown in the 'survival plot' of Fig. 4. The particle nearest the 'dynamic aperture' was found to stay in the storage ring for one million turns seemingly happily but getting lost in the very last 30 turns, as shown in Fig. 5.

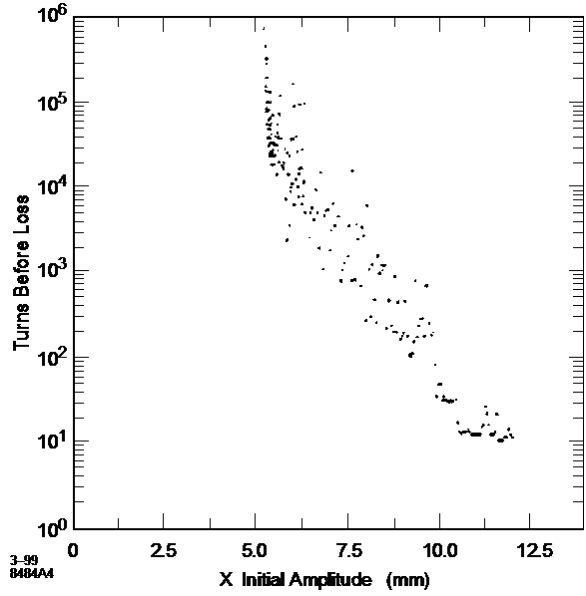

**Fig. 4:** A 'survival plot' to explore the dynamic aperture for the SSC by computer simulation. This result indicates a possible dynamic aperture in the neighbourhood of 5 mm. The last particle survived for $10^6$ turns.

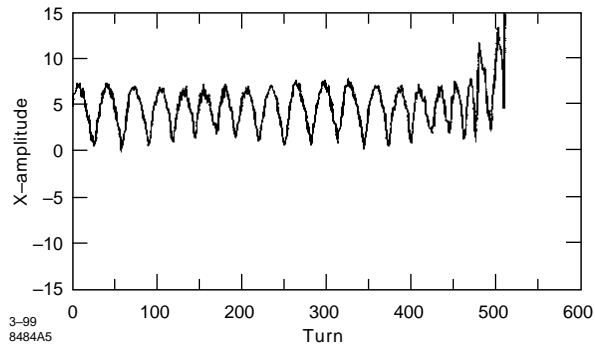

**Fig. 5:** The *x*-amplitude of the last surviving particle in Fig. 4 is shown for the very last 500 turns of its life. It shows that the particle was lost in a hurry in the very last 30 turns while showing no sign of unhappiness in its entire life before then.

What do we learn from this example? We learn that the particle loss mechanism is very subtle and that dynamic aperture prediction is difficult. In addition, it also says that it is impossible to 'predict' the stability of a particle by its 'long-term behaviour'. A hope to predict the $10^6$ turn stability of a particle by simulating, say, $10^4$ turns and hopefully detecting an 'early warning' signal, as was being pursued at the SSC, will not likely be fulfilled.

## 11. Detuned integrable systems

Accelerator designs necessarily start initially with an integrable system. So far, this initial integrable system has been chosen almost exclusively to be the linear, uncoupled Courant–Snyder system.

Our job then is to maximize the stability region when various perturbations are added to this initial system and thereby break its integrability. For this reason, the initial system must be chosen to be robust. For example, the choice of working point must avoid lower-order resonances, etc. However, chaos occurs as soon as integrability is broken, thus endangering the dynamic aperture. Even an infinitesimal perturbation can limit the dynamic aperture for an initially integrable system.

Is there a way to find another initial integrable system that is more robust than the linear uncoupled Courant–Snyder system? This is being tested at Fermilab by the IOTA project [10]. If the initial system contains a sizable 'detuning', it is believed to allow the initial system to tolerate larger perturbations. The situation is illustrated in Fig. 6.

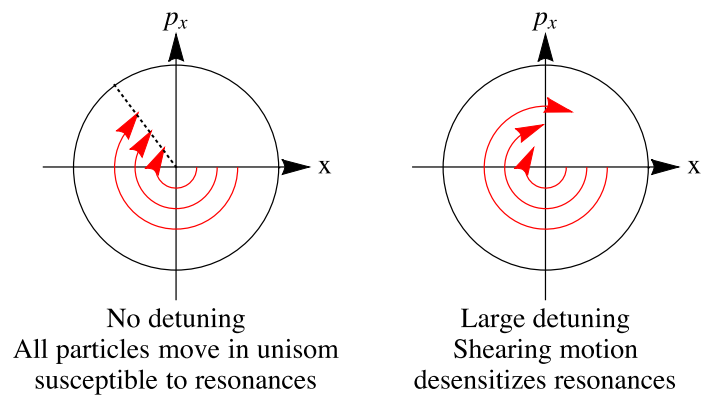

No detuning  
All particles move in unisom  
susceptible to resonances  

Large detuning  
Shearing motion  
desensitizes resonances  

**Fig. 6:** The initial integrable system can be a Courant–Snyder system (left-hand side) or a highly detuned system (right-hand side). The IOTA project aims to compare the robustness of these two systems.

The situation can be viewed as follows. In a conventional Courant–Snyder accelerator, both detuning (the stability mechanism) and chaos (the instability mechanism) originate from nonlinear perturbations $\varepsilon$. This necessarily means that both the detuning and the chaos are proportional to $\varepsilon$, i.e., they are both small and they fight each other with comparable strengths. There is no assurance who could have the upper hand.

In IOTA, by comparison, the detuning originates from a zeroth-order design. This means the detuning can be much greater than the chaos, thus in principle making the system much more robust. If proven correct, this means that what we have been doing in the past 70 years has not been wise!

## 12. Microbunches

One particular area deserving serious attention is the physics of microbunches. Microbunches are a new development, particularly brought to focus due to the advent of the free electron lasers as well as other new microbunching mechanisms. This will be a focus area of the beam dynamics of accelerator physics for many years to come. That microbunches are a valid subject is a direct consequence of the Liouville theorem.

The challenge of the microbunching beam dynamics is the dynamic range of the physics involved. Take the X-ray Free Electron Laser (FEL) for example. The electron bunch length is ~ 1 mm. The microstructures within the electron bunch are ~ 0.1 nm. So the dynamic range is $10^7$.

This large a dynamic range is very difficult to handle, both in simulations and in analysis. There is no way to simulate the beam dynamics in great detail as we did for single-particle models. To address the physics of microbunches, we invented yet another technique, frequency filtering, i.e., we only focus on the 'time evolution of one particular frequency component', with wavelength
$$\lambda_0 = \lambda_u (1 + K^2) / 2 \gamma^2 ,$$
of the beam distribution and Electromagnetic (EM) field distribution.

In particular, this single-frequency model of FEL physics does not contain information at any wavelength $< \lambda_0$. Results on the behaviour of the beam distribution become questionable when the beam bunch is shorter than $\lambda_0$. In conventional storage rings, this is called beam loading and the Robinson instability. For FELs, however, currently there are no simulation codes to deal with such cases.

Thus, for the FEL, we now have a peculiar beam dynamics model that is a mixture of the time domain and frequency domain. Such a mixture has always been considered an error-prone region to enter and has mostly been avoided in the past. The GENESIS code [11], a good mixture code, for example, for FEL simulations must be used with extreme care.

Incidentally, frequency filtering is a key ingredient in FEL physics not only in simulations. The analysis of FEL physics also filters out the one frequency component, as most readily evidenced by the famous cubic FEL equation or the SASE mechanism [12, 13].

## 13. Summary 1: the present landscape

The present landscape might look like Fig. 7.

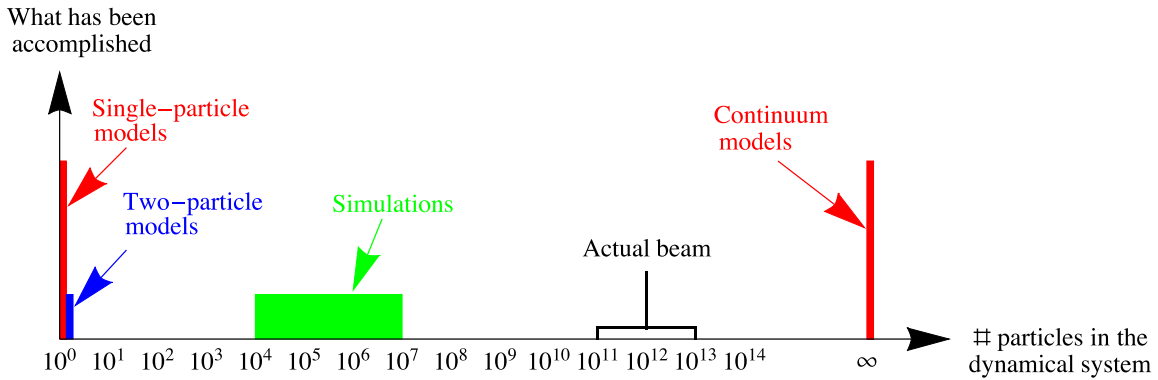

**Fig. 7:** A summary chart of our systems and what they have accomplished

Our approach so far has been to represent the real beam just as George Gamow described:

- a collection of non-interacting point particles – 1;
- two interacting particles + analysis – 2;

- one million interacting particles + simulation – 3;
- a continuum of phase space distribution – ∞;
- supplementing by binary collision models.

## 14. Summary 2: possible prospects in the future

As one tries to look a bit into the future, it can be expected that the industry standard of a joined force of TPSA and Lie algebra will continue. This then closes the door to the canonical perturbation theories, as is already happening. It is hoped that a purified algorithm based on SLIM and TPSA will be developed in some manner, thus avoiding the introduction of the Courant–Snyder special functions.

In addition, new developments are opening new doors. The FEL and other new microbunch mechanisms will blossom upon opening the door to mixed time- and frequency-domain approaches. The IOTA project might open a door to detuned systems. There is never a lack of excitement looking ahead from the viewpoint of accelerator physics. These future prospects are illustrated in Fig. 8.

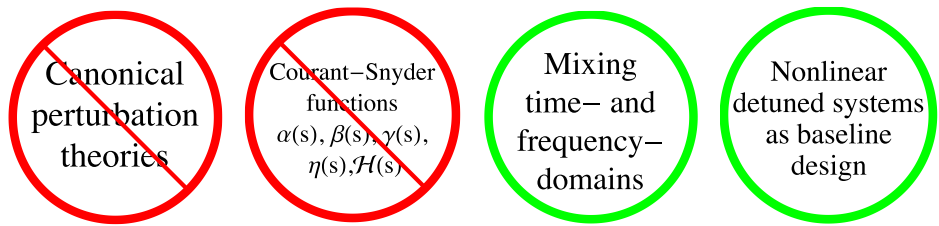

**Fig. 8:** A cartoonist view of possible future prospects of dynamical systems in accelerator physics

**Acknowledgement**
This work was supported by U.S. DOE Contract No. DE-AC02-76SF00515.